# Harnessing the Power of Hugging Face Transformers for Predicting Mental Health Disorders in Social Networks

**Alireza Pourkeyvan[1], Ramin Safa[1,*], Ali Sorourkhah[2]**

[1]Department of Computer Engineering, Ayandegan Institute of Higher Education, Tonekabon, Iran.
[2]Department of Management, Ayandegan Institute of Higher Education, Tonekabon, Iran.

## Abstract

Early diagnosis of mental disorders and intervention can facilitate the prevention of severe injuries and the improvement of treatment results. Using social media and pre-trained language models, this study explores how user-generated data can be used to predict mental disorder symptoms. Our study compares four different BERT models of Hugging Face with standard machine learning techniques used in automatic depression diagnosis in recent literature. The results show that new models outperform the previous approach with an accuracy rate of up to 97%. Analyzing the results while complementing past findings, we find that even tiny amounts of data (Like users' bio descriptions) have the potential to predict mental disorders. We conclude that social media data is an excellent source of mental health screening, and pre-trained models can effectively automate this critical task.

## 1. Introduction

There is no doubt that mental health is of vital importance to the health of an individual. Mental disorders increase the risk of developing other diseases, and without strong mental health, physical health would be impossible [1]. According to research by the World Health Organization (WHO), in 2015, more than 300 million people worldwide suffer from mental disorders. In addition to being the leading cause of disability worldwide, depression is one of the leading causes of death and suicide, killing approximately 800,000 people annually [2]. About 194 million people have been affected by Covid-19 since July 26, 2021. Due to the outbreak of this virus in 2019, people experienced a new lifestyle. Research has demonstrated that people infected with Covid-19 also exhibit symptoms of psychological disorders such as depression [3].

Despite the availability of various approaches for diagnosing and treating depression, the diagnostic process is subject to several limitations. These include insufficient funding for mental health initiatives, unequal access to healthcare services, inadequate numbers of mental health specialists, a lack of consultations with qualified psychologists or psychiatrists, social

Correspond Author: *safa@aihe.ac.ir*

stigma, and a significant misdiagnosis issue. The latter factor, namely improper diagnosis, is particularly noteworthy given its potential to impede effective treatment and exacerbate the condition [1], [4].

On the other hand, in recent years, there has been a significant increase in the utilization of social media, and an ever-growing number of individuals join various social platforms daily. Social media has revolutionized how people communicate, share information, and connect. Social media has enabled individuals and communities to easily share and create content and spread news, opinions, and ideas creating a new era of digital communication. Various fields, including marketing, politics, and keeping contact with friends and family, are affected by the power of social media to spread information [5], [6]. As a result of social media platforms, they can provide valuable information about people's mental health and how they interact with others socially and professionally [7].

Numerous social network platforms can be used for data analysis, including Twitter, Facebook, Instagram, and LinkedIn. In this paper, we propose an approach to predicting depression using Twitter data. Millions of active Twitter users make it one of the most popular social media platforms. As a result of its launch in 2006, Twitter has become one of the most popular platforms for real-time news updates, political discussions, and social interaction. According to the latest available statistics, Twitter has more than 330 million monthly active users, making it one of the world's most influential social networks [6]. A machine learning system is constructed using advanced statistical and probabilistic methods to build strategies that can improve through experience. This is a handy tool for predicting mental health. Using it, many researchers are able to gather information from the data, provide personalized experiences, and develop automated intelligent systems. Support vector machines, random forests, and artificial neural networks are examples of standard methods used in the literature [8].

BERT stands for Bidirectional Encoder Representation from Transformer. It is the state-of-the-art embedding model published by Google. This model has made significant strides in the field of NLP by providing better results in many NLP tasks, including question answering, text generation, sentence classification, and much more. Among the significant reasons for the success of BERT is the fact that it is a context-based embedding model as opposed to other popular embedding models, such as word2vec [9]. In this paper, we aim to identify the most effective approach to predicting depression from social media data by comparing four different BERT models from Hugging Face with standard machine learning techniques used in automatic depression diagnosis in recent literature. Our results provide insights into the potential of social media data for predicting depression and have implications for early detection and intervention.

The rest of this paper is organized as follows. The next section reviews related work on predicting depression using social data. We discuss common approaches, traditional ways, and recent advances in deep learning models such as transformers. We also highlight the features studied in previous works and describe the evaluation metrics used to assess these models' performance. The third section describes our approach to predicting depression using Twitter data and transformers. We provide a detailed design explanation and the preprocessing steps required to prepare the data for analysis. The following section presents our experimental analysis. We describe the implementation details and present the results of our experiments. Next, we compare the performance of the four BERT models with the baseline and discuss the challenges we faced during the implementation phase. Finally, we discuss the implications of our results and provide insights into the potential of social data for predicting mental health. We conclude the paper with a summary of our findings and suggestions for future research.

## 2. Related work

Traditionally, these approaches have been based on handcrafted features and rule-based models. Standard methods include sentiment analysis, topic modeling, linguistic analysis/ psycholinguistic analysis, network analysis, and traditional machine learning approaches [10]. These tasks will require us to use machine-learning approaches, as shown in
Figure 1.
In the following paragraphs, we provide a brief overview of the relevant approaches in the literature; for more detailed information, we encourage the authors to read references [11] and [12].

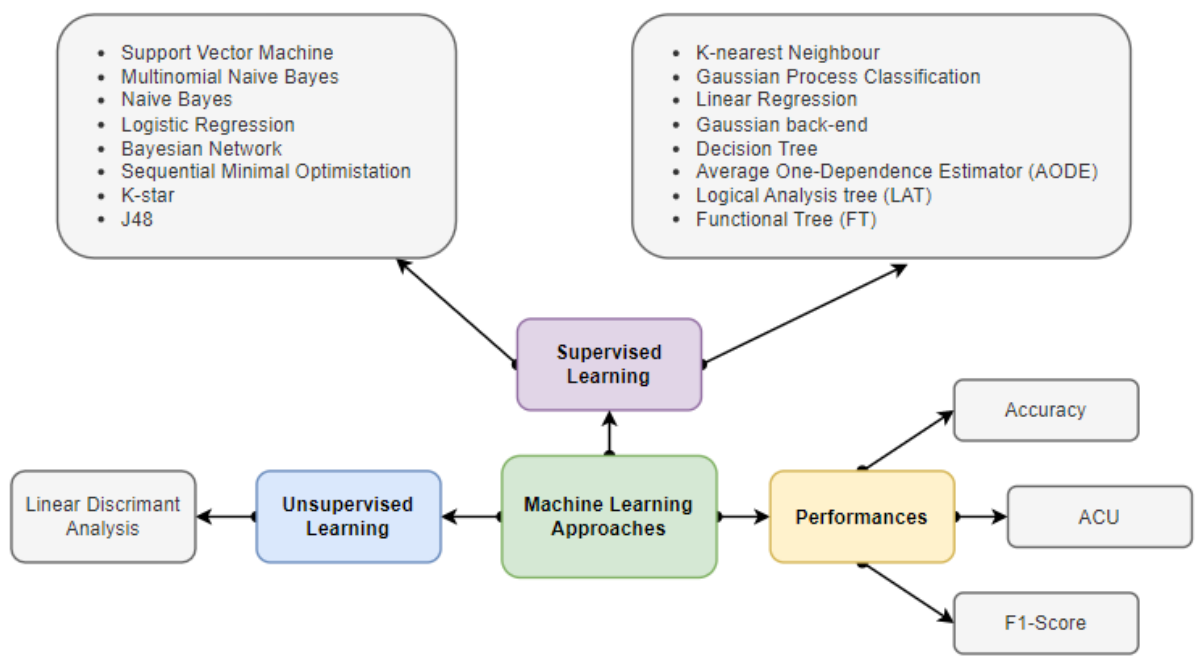

**Figure 1.** Traditional machine learning approaches [13].

Support Vector Machine (SVM) was developed as a popular classification technique. Classification tasks generally involve separating data into training and testing sets. In SVM, the goal is to produce a model based on the training data which predicts, based solely on the test data attributes, the target values of the test data. K-Nearest Neighbor (KNN) is a simple algorithm that stores all the available cases and then classifies new cases based on a similarity measure (e.g., distance functions). KNN has been used in statistical estimation and pattern recognition as a non-parametric technique. Cases are classified based on the majority vote of their neighbors and are assigned to the class most commonly shared by their K nearest neighbors. If K = 1, then the case is simply assigned to the class of its nearest neighbor. The random forest consists of a combination of tree predictors in which every tree is determined by the values of a random vector sampled independently and with the same distribution for all trees within the forest [14]. The Multilayer Perceptron (MLP) is also a type of Artificial Neural Network (ANN) designed to accurately map a given set of inputs to corresponding outputs. It achieves this by combining multiple layers of perceptron, which are mathematical functions, into a single complex function. Each perceptron takes one or more inputs, applies a set of weights and biases to them, and produces an output. The output of each perceptron is then passed through an activation function, which introduces non-linearity into the model, and the resulting output is used as input for the next layer [15]. Its versatility and ability to learn complex relationships make it a popular choice for various applications in various classification fields. It should be noted that most of these approaches have limitations regarding the ability to capture the complex and nuanced language used in social media posts, which is influenced by cultural and contextual factors. A recent development in deep learning models, such as transformers, has enabled researchers to achieve state-of-the-art results in predicting mental health conditions [16].

Transformers are based entirely on a mechanism of attention known as self-attention. An encoder-decoder architecture makes up the transformer. We feed the input sentence to the encoder, which learns the representation of the input sentence and sends the representation to the decoder. The decoder receives input from the encoder, and output is produced. The encoder of the transformer is bidirectional, meaning it can read a sentence both ways. We can perceive BERT as the transformer, but only with the encoder [9]. Recent developments in transformers have shown promising results when detecting mental health conditions on social media platforms. This architecture has been demonstrated to achieve state-of-the-art performance on various natural language processing (NLP) tasks and can capture contextual information. It is also possible to fine-tune transformer models, such as BERT, for specific tasks. Recent studies have demonstrated that BERT effectively captures contextual information and performs at a state-of-the-art level on several NLP tasks. BERT can be fine-tuned for specific tasks, such as depression prediction based on social media data.

Much research has demonstrated that BERT performs better than traditional approaches in detecting mental health conditions on social media platforms [17]. MentalBERT was developed based on posts containing mental health-related information collected from Reddit and is based on BERT-Base. This model follows the standard pretraining protocols for BERT and RoBERTa, and is trained and released using Hugging Face's Transformers library to facilitate the automatic identification of mental disorders in online social content using masked language models [18].

Table 1 reviews previous research on depression detection on social media platforms, focusing on traditional approaches such as binary classification and current methods using pre-trained transformer models such as BERT.

**Table 1.** Recent related studies

| Author, date, reference | Mental health issue | Dataset | Platform | Machine learning method(s) |
|---|---|---|---|---|
| The implemented strategy | Depression | Autodep (Textual analysis) | Twitter | DisitlBERT, BERT, MentalBERT and DistilRoBERTa |

| Safa et al. (2023) [19] | Depression | Autodep (Multimodal analysis) | Twitter | Decision Tree, Linear SVM, Gradient Boosting Classifier, Random Forest, RidgeClassifer, AdaBoost, Catboost, and MLP |
| --- | --- | --- | --- | --- |
| Kabir et al. (2023) [20] | Depression | Tweets | Twitter | BERT, DistilBERT |
| Ilias et al. (2023) [21] | Stress, Depression | Public datasets | Reddit | MentalBERT |
| Devika et al. (2023) [22] | Depression, Suicide | Posts | Reddit | BI-LSTM, BERT |
| Triantafyllopoulos et al. (2023) [23] | Depression | Pirina dataset | Reddit | BERT |
| Benítez-Andrades et al. (2022) [24] | Eating disorder | Tweets | Twitter | BERT |
| Zeberga et al. (2022) [4] | Depression, Anxiety | Posts | Twitter, Reddit | BI-LSTM, BERT |
| Nisa et al. (2021) [25] | Depression, Self-harm | Comments | Reddit | BERT |
| Bucur et al. (2021) [26] | Gambling, Self-harm, Depression | Posts | Reddit | BERT |
| Singh et al. (2021) [27] | COVID-19 impact on social life | Tweets | Twitter | BERT |
| Sekulic et al. (2020) [28] | Mental health | SMHD dataset | Reddit | Logistic Regression, SVM |
| Cacheda et al. (2019) [29] | Depression | Posts and comments | Reddit | Random Forest |
| Islam et al. (2018) [30] | Depression | Comments were collected with NCapture | Facebook | KNN |
| Peng et al. (2017) [31] | Depression | User microblog text, user profile, and user behaviors | Twitter | SVM |

In conclusion, our review of previous research on depression detection on social media platforms shows that traditional approaches, such as binary classification, have achieved moderate performance. In contrast, new methods using pre-trained transformer models, such as BERT, have been found to have achieved higher performance.

We will discuss the effectiveness of using pre-trained BERT models from the Hugging Face library to detect mental health issues on social media in the following section, demonstrating that this approach can achieve high accuracy without requiring the creation of new models. By using this approach, we will be able to achieve high performance using fewer computational resources.

## 3. Methodology

The use of social media in mental health research has gained popularity over the past few years since it provides a rich source of information about users' thoughts, feelings, and behaviors. The potential of social media in mental health research is vast. As previously stated, researchers use it to gain insights into the mental health of a population, track changes in mental health over time, and identify risk and protective factors for mental health issues [32]. Automated systems for detecting mental health issues have been developed to analyze textual data, with NLP and transformer models being utilized to identify patterns [7].

An effective method for diagnosing mental health disorders in social networks involves analyzing users' self-report statements. This approach has shown promise in detecting mental health symptoms. It can be used to collect positive and negative samples, which can then be automatically validated to train a machine-learning model. Previous research has shown that this method can efficiently detect mental health symptoms. However, the potential impact of BERT models, which have not yet been explored in this context, remains unknown. Given the success of these models in various fields, we aim to investigate their effectiveness in the domain of mental health diagnosis in the social data [19], [33].

As part of this study, we propose using Twitter users' tweets and bios to predict depression. Specifically, we use BERT models from the Hugging Face library, which have been fine-tuned based on large datasets of reviews, tweets, and other textual data. A transformer-based deep learning model has demonstrated impressive results in various NLP tasks, such as sentiment analysis, question answering, and text classification [9]. The Hugging Face open-source transformers library in the NLP community is trendy and useful for several Natural Language Understanding (NLU) tasks. The library contains thousands of models that have been pre-trained in more than 100 languages.

Using pre-trained BERT models from the Hugging Face library, we can predict symptoms based on users' textual data.

Figure **2** presents a four-step approach that involves data selection, preprocessing, training, and validation modules. To determine which model is best suited to our problem, we tested four BERT models, distilbert-base-uncased-finetuned-sst-2-english (DBUFS2E) [34], bert-base-uncased (BBU) [35], mental-bert-base-uncased (MBBU) [18] and distilroberta-base (DRB) [36]. The following parts describe the methodology in detail, covering fine-tuning and evaluation. Eventually, we analyze the performances of the four models and explore the potential implications for mental health research.

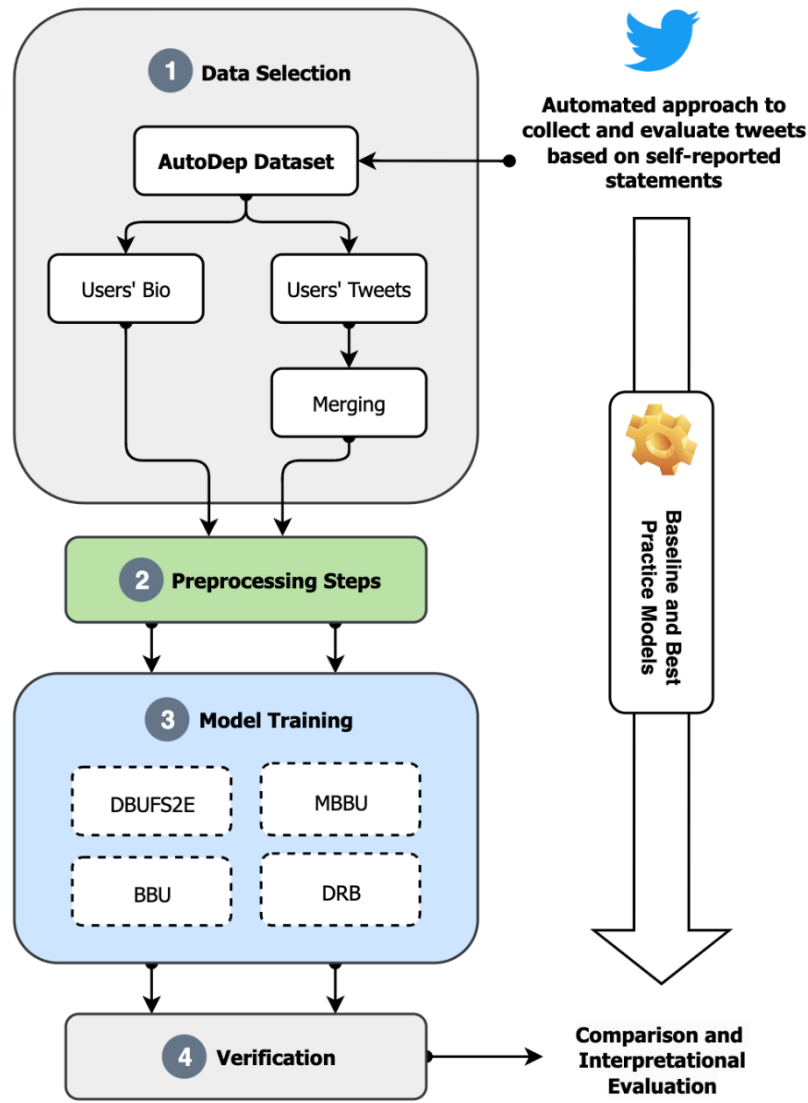

**Figure 2.** Data Analysis Framework

### 3.1. Data selection

Data collection is the process of gathering and analyzing information on targeted variables in a systematic manner that enables one to answer stated research questions, test hypotheses, and evaluate results. For the purpose of training models, the collection of data is a critical component in providing the necessary inputs to create models capable of accurately predicting outcomes.

The present study utilizes the Autodep dataset[1], which was automatically collected and evaluated through the Twitter API [19]. It encompasses a range of data, including posts, bio descriptions, profile pictures, and banner images of Twitter users who have publicly disclosed their mental health status. To ensure the authenticity of the results, benchmarking techniques were employed to compare the outcomes of various measures. The dataset contains 11,890,632 tweets and 553 bio-descriptions. In the last study on predictive models for depressive symptoms on the Autodep, utilizing only tweets and bio-texts resulted in accuracy rates of 91% and 83%, respectively [19]. Here we'll extract the user bios and tweets from the raw data and save them as separate files. To analyze the tweets holistically, all tweets for each user were merged into a single cell. The following section will detail our actions to prepare the data for analysis.

---
[1] https://github.com/rsafa/autodep

### 3.2. Preprocessing Steps

As part of the preparation process for training our model, data preprocessing is an essential step. Social media posts with hashtags, links, special characters, and emojis can often be noisy. However, these elements may not add much value to the text. We used several preprocessing steps to clean and extract meaningful information from the studied features (tweets and bio descriptions).

---
**Algorithm 1.** Data Preprocessing Steps

---
**Input:**
    Tweets and Bio descriptions (control and diagnosed group)
**Output:**
    Preprocessed data
**Method:**
    **for each** value in groups
        **if** the value is not in English, skip the value
        **if** the value starts with "RT", skip the value
        **if** the value starts with "@", skip the value
        **if** the value contains URLs, remove them from the value
        **if** the value contains emoji, remove them from the value
        **if** the value contains any special character, sanitize the value
        convert the value to lowercase
        **if** the value contains any additional spaces, remove them
        tokenize the value
        **if** the value contains stop words, remove them
        lemmatize values
    **end for**

---

As illustrated in Algorithm 1, the first step is to check the language, retweets, and mentions. Due to the fact that they do not add value, they should be skipped. Next, URLs will be eliminated from the text utilizing regular expressions, as URLs are improbable to be associated with the prognosis of mental disorders. URLs have a pattern that begins with http/https or www, and we will use regular expressions to match and remove URLs from the text.

Next, we will remove all emojis from the text since they may not provide significant contextual information. We will remove the emojis since we wish to keep our approach as straightforward as possible. Emojis follow an ASCII pattern that can be deleted using a regular expression. We will also remove "*", "^", "@", etc., using regular expressions to improve noise reduction and a smaller dataset. As part of the removal process, we will make sure to eliminate any extra spaces that might remain.

Following, the sentence will be converted to lowercase to ensure consistency. As text data can have different case types, the conversion to lowercase will ensure that the same word will be treated as identical regardless of its case. To enhance the quality of the text, we shall remove any extra spaces between tweets that may impact the algorithm's performance.

After that, we will tokenize the text using the NLTK library, which will transform long texts into smaller units known as tokens. Tokenization is the process of splitting the text into individual words or tokens that can then be used as input for further analysis.

We will perform using the Natural Language Toolkit NLTK [37] library to eliminate insignificant and meaningless stop words from the English language, such as 'I', 'am', 'a', 'the', 'of', 'to', etc. And we will apply lemmatization to group together different inflected forms of words to allow them to be analyzed together as a single item. To reduce the complexity of text data, lemmatization is a process of normalizing words into their base or root form, which reduces the number of variables. A lemmatization will be performed using the WordNetLemmatizer[1] from the NLTK library. We will also be conducting experiments with stemming, which reduces words to their basic form by eliminating suffixes from their root form. However, we found that lemmatization provided slightly better results than stemming, with a minimal difference between the two.

After completing the preprocessing steps, we will verify that the length of the character does not fall below five characters to remove any unnecessary words or characters that may have been left over from the previous steps. Then, we will return the results to the training section.

Figure **3** provides an example of the aforementioned preprocessing steps as follows.

### 3.3. Model training

Upon completion of the preprocessing steps, we will utilize the Hugging Face library to fine-tune four pre-trained BERT models: distilbert-base-uncased-finetuned-sst-2-english, bert-base-uncased, distilroberta-base, and mental-bert-base-uncased. This approach will enable us to develop a predictive model for detecting depression.

Distilbert-base-uncased-finetuned-sst-2-english is a text classification model developed by Hugging Face that uses distilbert-base-uncased for topic classification, which was trained on the SST-2 dataset. In addition to fine-tuning downstream tasks, the model can also be used to model masked language or predict the next sentences. The distilBERT model has been developed to simplify the original BERT model, making it smaller, faster, and more efficient while maintaining most of its performance.

The bert-base-uncased model is a pretrained NLP model that has been introduced in a paper. This model has been trained on a large corpus of English data using a self-supervised approach; in other words, it has been trained on raw texts without human labeling. A model was trained on two objectives: Masked Language Modelling (MLM) and Next Sentence Prediction (NSP). MLM involves randomly masking 15% of the words in a sentence and predicting the masked words, allowing the model to learn a bidirectional sentence representation. In NSP, two masked sentences are concatenated, and a prediction is made about whether the two sentences follow each other or not. The learned features of the BERT model can be applied to downstream tasks, such as classification.

---

[1] https://www.nltk.org/_modules/nltk/stem/wordnet.html

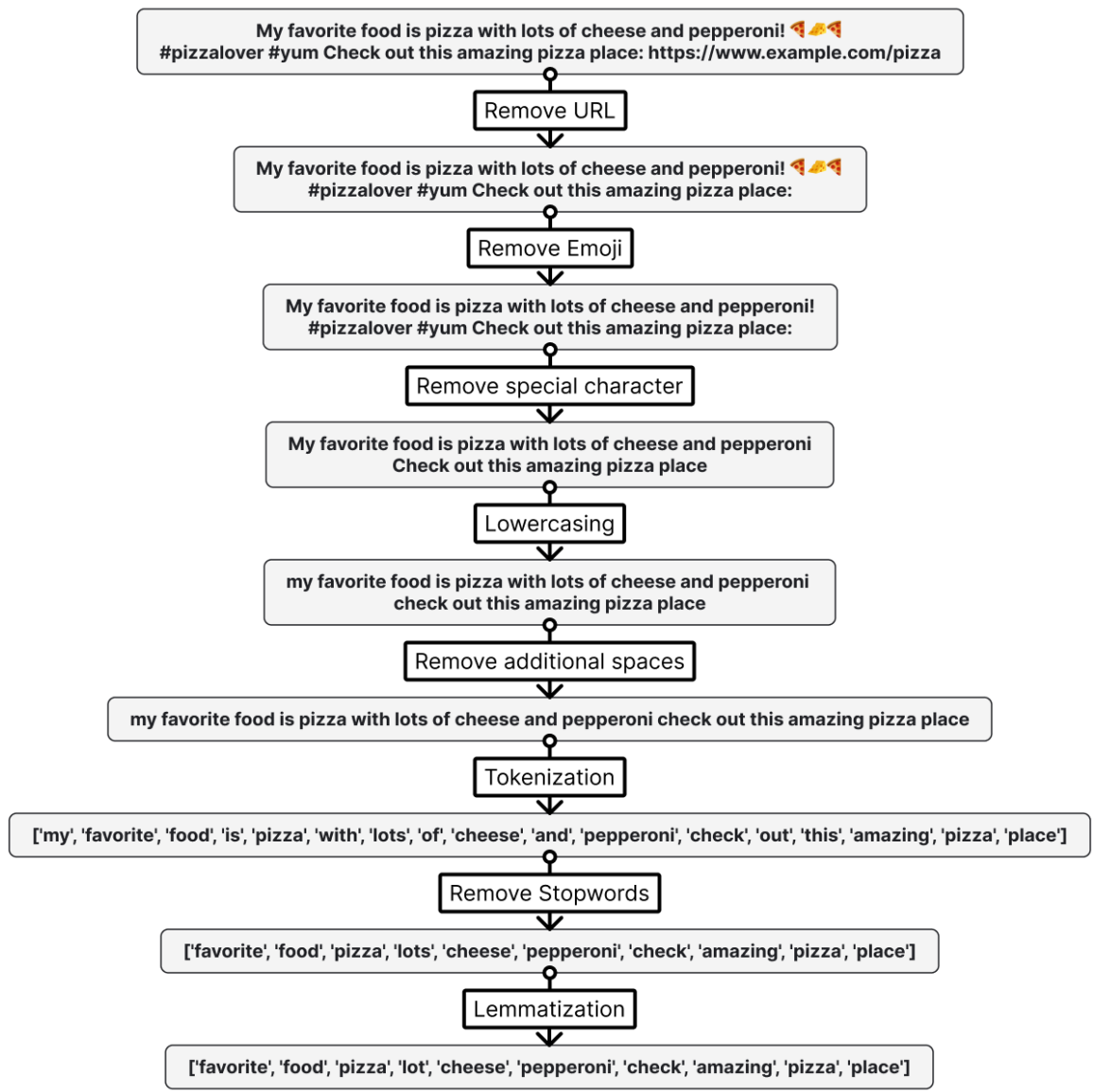

**Figure 3.** Preprocessing steps

The Roberta-base model was pre-trained on a large corpus of English data using the MLM objective without human labeling. The model can learn a bidirectional representation of the sentence to extract useful features for downstream tasks, such as sequence classification, token classification, or question answering. It is case-sensitive, meaning it differentiates between different capitalizations. There is a distilled version of the Roberta-base model, distilroberta-base, which follows the same training procedure as DistilBERT [38]. Compared to the Roberta-base model, it has fewer layers, dimensions, and parameters, resulting in a smaller, faster, and more efficient model while maintaining its performance in most cases.

As discussed earlier, MentalBERT is a variant of the bert-base-uncased model that has been fine-tuned on a dataset of mental health-related posts from Reddit. This allows the model to capture better the nuances and complexities of mental health language, which can be useful for tasks such as sentiment analysis, classification, or question-answering on related content. The learned features of the BERT model can be applied to downstream tasks, and the use of MentalBERT represents an important step forward in using NLP to address mental health issues and improve our understanding of this important area.

We found that utilizing pre-trained models gave our approach a significant advantage. By using the pre-existing weights and architectures, we were able to decrease the amount of computational resources needed for training while also enhancing the model's performance. As a result of their training on large quantities of text data, these models could learn complex patterns and relationships within the text, which is crucial for detecting depression.

As a part of our methodology, we loaded the datasets as CSV files and divided them into training and testing sets using an 80/20 split. It is worth noting that two distinct datasets were used to apply this split: tweets and bios. As a next step, we applied tokenization to the text data using the pre-trained BERT tokenizer that was specific to the model we were fine-tuning. These tokens are then utilized as inputs in the model for further analysis and processing.

As an integral component of the training process, we partitioned the datasets into ten distinct folds and shuffled them before initiating the training process. Using K-fold cross-validation, we avoided overfitting and improved model accuracy. Each fold was trained separately, using five epochs per training. The model was trained on the training set and evaluated on the validation set at each training epoch. We loaded, preprocessed, and fine-tuned pre-trained BERT models for both tweets and bios using the Hugging Face library and the Trainer method. As a result of fine-tuning the pre-trained models with our preprocessed Twitter dataset, we could construct high-performing depression detection models. The K-fold cross-validation technique ensures that our model is trained and evaluated on a diverse dataset, thus ensuring quality and accuracy.

### 3.4. Validation

Several metrics were used to evaluate the performance of our fine-tuned BERT models for depression detection, including accuracy, F1 score, receiver operating characteristic (ROC), and the area under the curve (AUC).

An accuracy measure indicates how many instances were correctly predicted. The F1 score measures the balance between precision and recall and represents the harmonic mean of precision and recall. Precision measures the proportion of true positive predictions among all positive predictions made by the model. Recall, on the other hand, measures the proportion of true positive predictions among all actual positive instances in the data. AUC indicates how well the model performed regarding true positive rates (TPRs) and false positive rates (FPRs). An AUC represents the area under a ROC curve, which plots the TPR against the FPR. Using the confusion matrix derived from the predictions of the models, the accuracy and F1 score were calculated using Equations 1 to 4 for each fold:

$$Accuracy = \frac{(TP + TN)}{(TP + TN + FP + FN)} \quad (1)$$

To calculate the F1 score, we first compute the precision and recall using Equation 2 and Equation 3.

$$Precision = \frac{TP}{(TP + FP)} \quad (2)$$

$$Recall = \frac{TP}{(TP + FN)} \quad (3)$$

$$F1\ score = \frac{2\ (precision \times recall)}{(precision + recall)} \quad (4)$$

TP (True Positive) represents the number of correctly predicted positive instances, TN (True Negative) represents the number of correctly predicted negative instances, FP (False Positive) represents the number of incorrectly predicted positive instances, and FN (False Negative) represents the number of incorrectly predicted negative instances.

For the final evaluation of the performance of the models, we used K-fold cross-validation. We split the dataset into ten folds and trained each fold separately, using five epochs per training. At each epoch, the model was trained on the training set and evaluated on the validation set.

To assess the final performance of the models, we employed K-fold cross-validation. Specifically, we partitioned the dataset into ten folds and trained each fold individually, with a training duration of five epochs. At each epoch, the model was trained on the training set and evaluated on the validation set. This process was repeated for each of the ten folds, resulting in a comprehensive evaluation of the model's performance across the entire dataset. By utilizing K-fold cross-validation, we aim to provide a robust and reliable assessment of the model's performance, while minimizing the risk of overfitting to a specific subset of the data.

## 4. Experimental analysis

We discussed various concepts that provide the foundation for our methodology for predicting depression in the previous sections. Throughout this section, we examine experimental analysis in greater depth to assess our approach's effectiveness.

Our implementation was carried out using Python programming language and the Google Colab platform, which enabled us to leverage the computing power of a T4 GPU. We utilized two distinct datasets comprising users' tweets and bios to fine-tune the pre-trained BERT models. This process involved adapting the pre-existing models to our data's specific domain, which helped optimize their performance on our task. By fine-tuning the BERT models on these datasets, we could tailor their representations to capture the unique characteristics of our data and achieve state-of-the-art results.

For the purpose of comparing the performance of each model and evaluating their precision and recall for detecting depression, we also plotted the confusion matrix for the entire model. Figure 4 shows the confusion matrixes for the entire models, DBUFS2E, BBU, MBBU, and DRB, derived from the tweet and bio datasets. Based on the confusion matrix, it was evident that almost all models were highly accurate and reliable in detecting depression, with only a small number of false positives and false negatives.

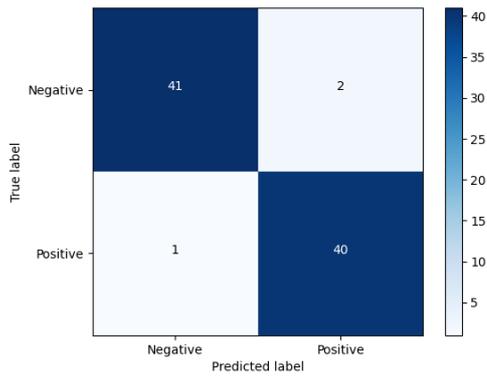

a) Bios with DBUFS2E

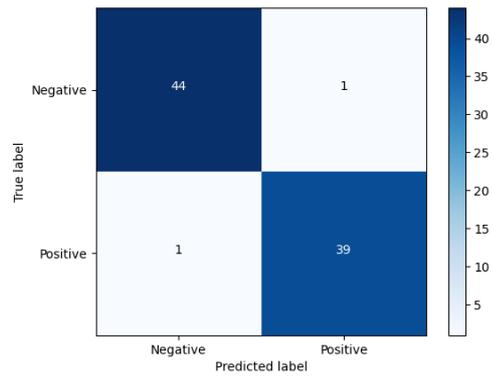

b) Tweets with DBUFS2E

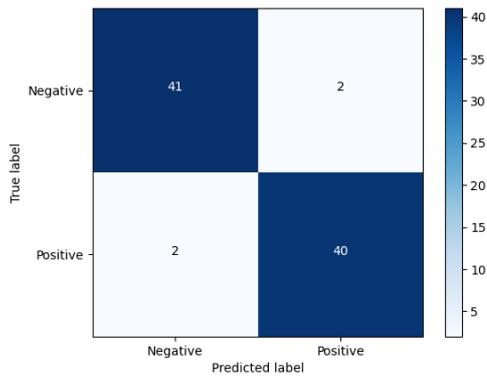

c) Bios with BBU

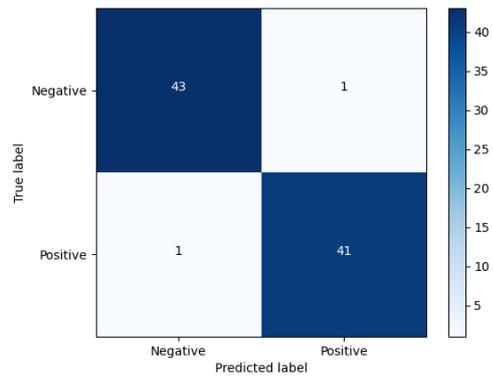

d) Tweets with BBU

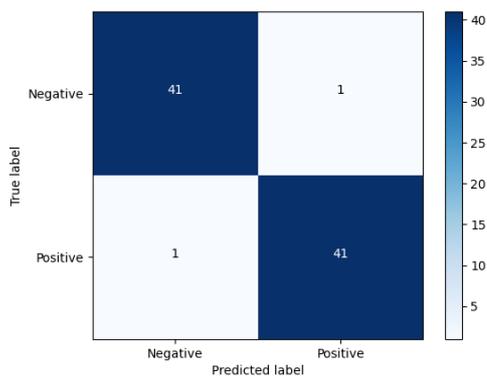

e) Bios with MBBU

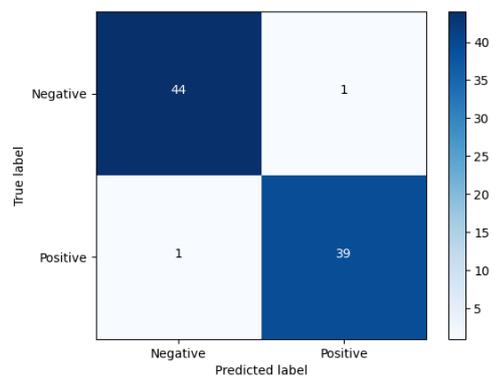

f) Tweets with MBBU

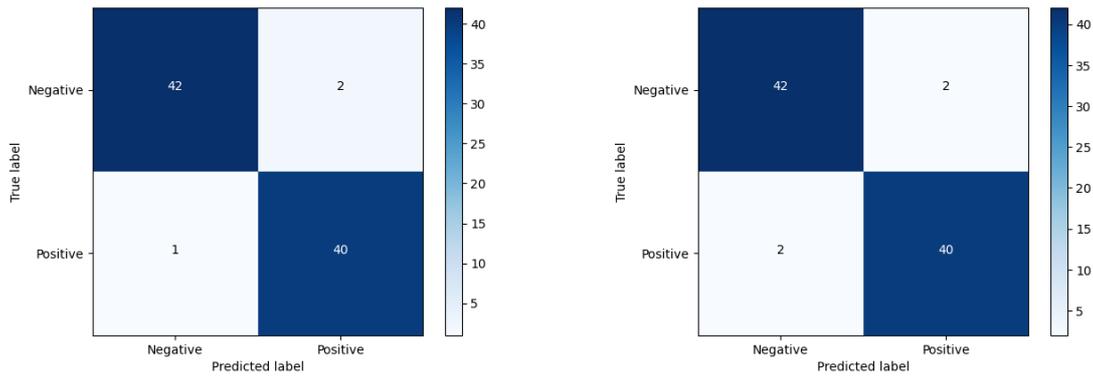

g) Bios with DRB    h) Tweets with DRB

**Figure 4.** Confusion matrixes for users' tweets and bio descriptions.

Our approach of fine-tuning pre-trained BERT models for detecting depression using Twitter data has yielded noteworthy results. The high levels of accuracy, F1 score, and AUC scores achieved for both tweets and bios are a testament to the efficacy of our method.

Table 2 presents the results of our analysis of various methods for predicting disorder on research data. To evaluate their effectiveness, we used Logistic Regression as the baseline classifier and compared its performance with DBUFS2E, BBU, MBBU, and DRB. Our selection of features for this analysis was based on prior research [19], which indicated that tweets word bigrams and bio-character four-grams were among the most accurate embedding methods for predicting disorder when used in conjunction with different classifiers. Likewise, to provide a comprehensive and unbiased assessment of our NLP models, we have incorporated MLP and Catboost into our evaluation. These models were also selected based on the findings of a prior investigation, which demonstrated their superior performance compared to other approaches for the task at hand. By including a range of models in our evaluation, we aim to provide a more complete picture of the strengths of each approach, which can help to inform future research in this area.

Based on the results presented in Tables 2 and 3, it is apparent that the BBU and MBBU models exhibit the highest metric values among all models for both the Tweet and Bio datasets. Notably, these models significantly outperform other standard methods that have been used in prior research. These findings corroborate previous studies demonstrating a strong association between user-generated content and mental health status. Especially, the short bio-text is shown to have a substantial impact on the detection of clues related to depressive disorder. These results underscore the potential of these models for early detection and prevention of mental health issues.

Figure 5 also shows the ROC curves for DBUFS2E, BBU, MBBU, and DRB models on the tweet and bio datasets. For both datasets, the DBUFS2E model achieved the highest mean AUC values, with the bio dataset achieving a mean AUC of 0.96 and the tweet dataset achieving a mean AUC of 0.98. Additionally, the BBU model performed well, with a mean AUC value of

0.95 and 0.97 for the bio and tweet sets. According to the ROC curves, all models achieved high TPRs and low FPRs, indicating that the models could accurately detect depression.

Table 2. Accuracy, F1 score, and AUC for tweets analysis

| Model | Accuracy | F1 score | AUC |
|---|---|---|---|
| DBUFS2E | 0.97 | 0.96 | 0.98 |
| BBU | 0.97 | 0.97 | 0.98 |
| MBBU | 0.96 | 0.96 | 0.98 |
| DRB | 0.95 | 0.95 | 0.97 |
| **Baseline** | 0.75 | 0.75 | 0.75 |
| Catboost | 0.91 | 0.89 | 0.91 |

Table 3. Accuracy, F1 score, and AUC for bios analysis

| Model | Accuracy | F1 score | AUC |
|---|---|---|---|
| DBUFS2E | 0.95 | 0.96 | 0.96 |
| BBU | 0.95 | 0.94 | 0.96 |
| MBBU | 0.96 | 0.96 | 0.96 |
| DRB | 0.95 | 0.95 | 0.96 |
| **Baseline** | 0.67 | 0.62 | 0.67 |
| MLP | 0.83 | 0.82 | 0.83 |

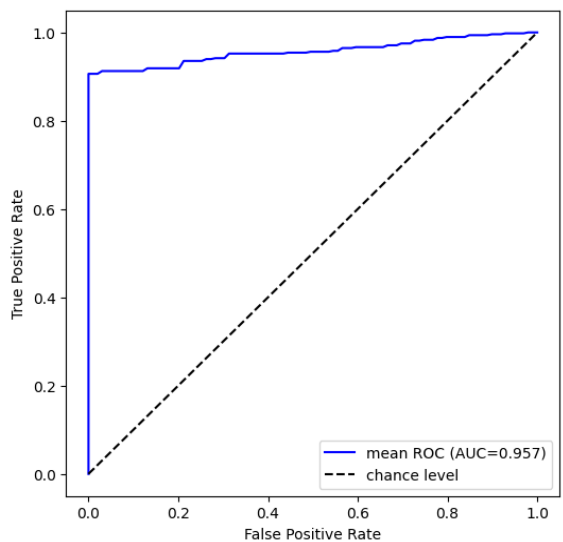
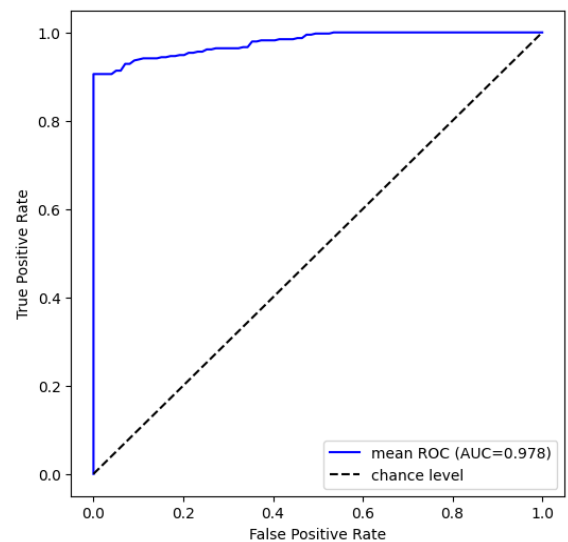

a) Bios with DBUFS2E                b) Tweets with DBUFS2E

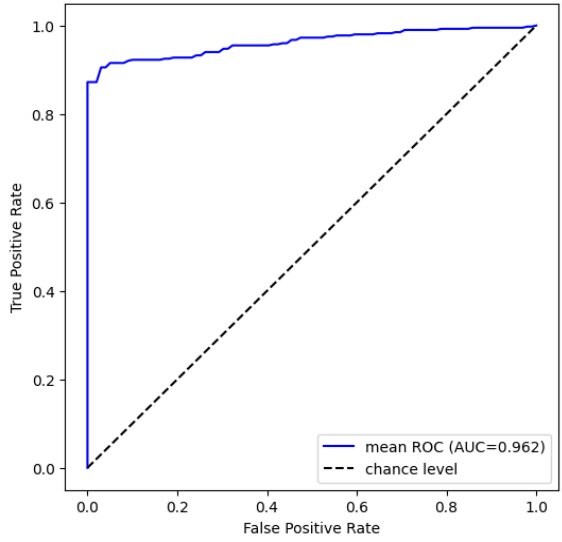 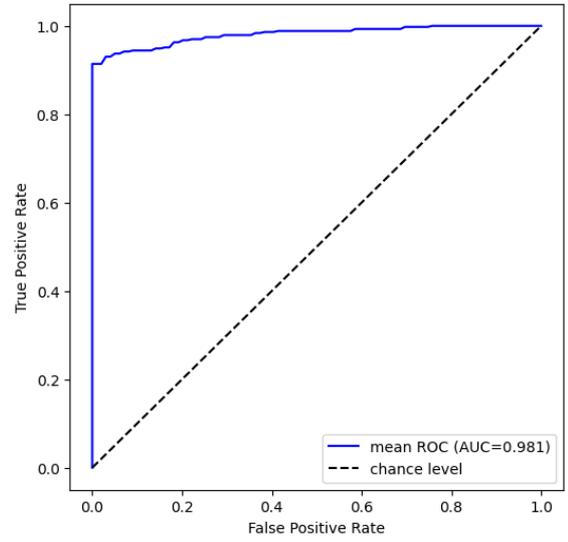

c) Bios with BBU      d) Tweets with BBU

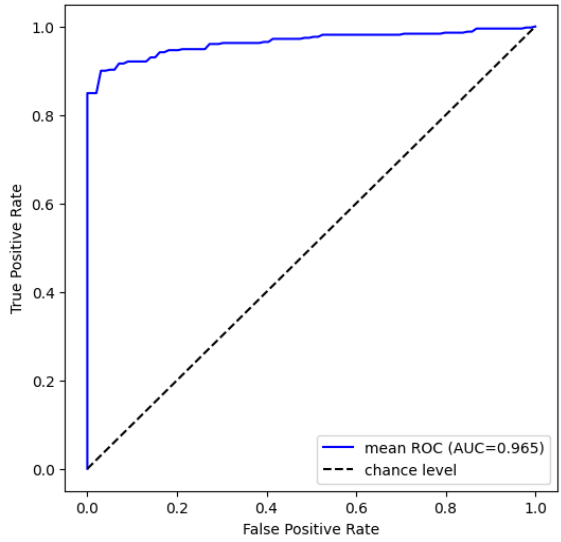 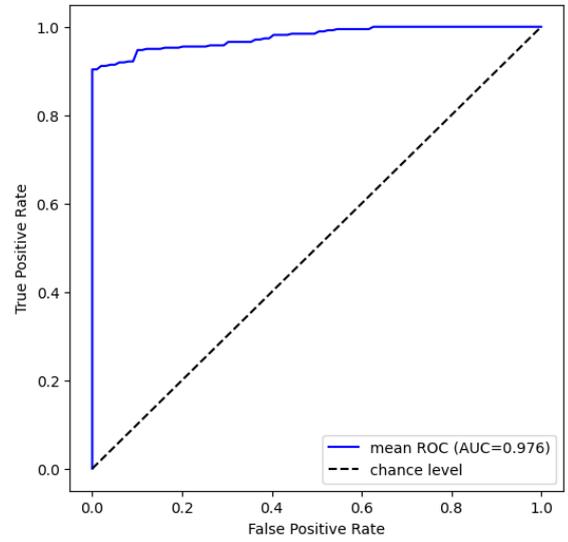

e) Bios with MBBU     f) Tweets with MBBU

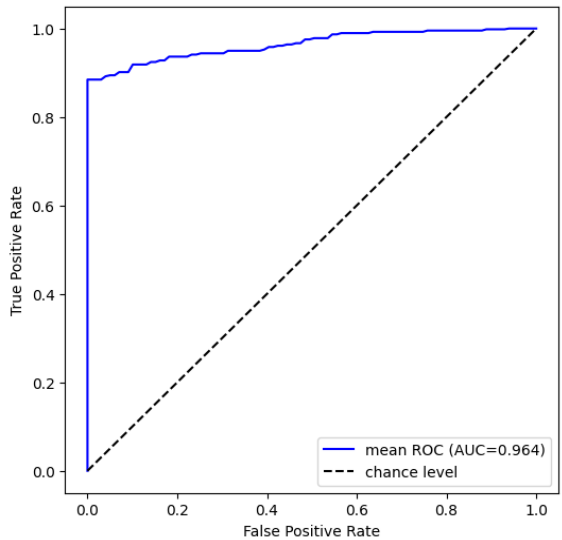 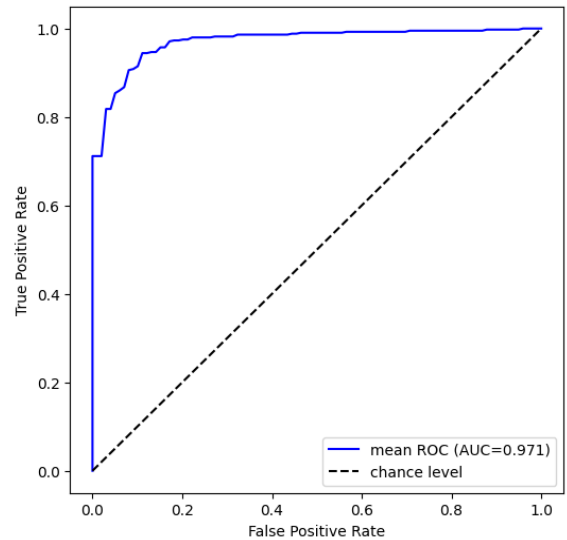

g) Bios with DRB          h) Tweets with DRB

**Figure 5.** ROC curves for bio and tweet

## 5. Conclusion and future work

Our study underscores the importance of analyzing social media data for mental health research, as it can provide valuable insight into an individual's mental health status. By examining Twitter data, we demonstrate the potential of using such information to predict depression symptoms with high accuracy and F1 scores compared to previous findings, which is a promising development in the automatic detection of depression symptoms. We analyzed tweets and bio information using four different BERT models from the Hugging Face and evaluated the models' performance using 10-fold cross-validation and standard measurements in the literature. The results show a more than 6% increase in metrics compared to previous research.

It should be noted that additional user data, including images, emojis, and hashtags, can enhance predictions. Future research can explore combining data types to improve the performance of the models; for instance, a more comprehensive view of mental health status may be obtained by combining tweets and bio information. Our approach may also be applied to other mental health conditions to improve early detection and intervention, such as anxiety and post-traumatic stress disorder. Our findings confirm that social data can be a valuable tool for mental health screening and monitoring and can assist in improving our understanding of mental health conditions and developing effective interventions. We hope this study inspires future research into social data mining for mental health research and ultimately improves mental health outcomes for individuals worldwide.